\documentclass[aps,prl,twocolumn,groupedaddress,nofootinbib,longbibliography]{revtex4-1}
\usepackage{amsmath, amsfonts, amssymb,amsthm}
\usepackage{graphicx,subfigure}
\usepackage{mathrsfs}
\usepackage[mathscr]{eucal}
\usepackage[T1]{fontenc}
\usepackage{physics}
\usepackage{bm}
\usepackage{subfigure}
\usepackage[bookmarks=false]{hyperref}
\usepackage{xcolor}
\hypersetup{colorlinks=true,citecolor=blue,
linkcolor=blue,urlcolor=blue,pdfstartview=FitH,
bookmarksopen=true}

\DeclareMathAlphabet{\mathpzc}{OT1}{pzc}{m}{it}

\begin{document}
\title{Two geometric phases can dramatically differ from each other even if their evolution paths are sufficiently close in a pointwise manner}
\author{Da-Jian Zhang}
\affiliation{Department of Physics, Shandong University, Jinan 250100, China}
\author{P. Z. Zhao}
\affiliation{Department of Physics, Shandong University, Jinan 250100, China}
\author{G. F. Xu}
\affiliation{Department of Physics, Shandong University, Jinan 250100, China}

\date{\today}

\begin{abstract}
One milestone in quantum physics is Berry's seminal work [Proc.~R.~Soc.~Lond.~A \textbf{392}, 45 (1984)], in which a quantal phase factor known as geometric phase was discovered to solely depend on the evolution path in state space. Here, we unveil that even an infinitesimal deviation of the initial state from the eigenstate of the initial Hamiltonian can yield a significant change of the geometric phase accompanying an adiabatic evolution. This leads to the surprising observation that two geometric phases can dramatically differ from each other even if their evolution paths are sufficiently close in a pointwise manner.

\end{abstract}

\maketitle

Ever since its publication in 1984, Berry's seminal work \cite{1984Berry45} has become increasingly influential throughout quantum physics. This work reports that apart from the familiar dynamical phase, the evolved state of a quantum system undergoing an adiabatic evolution acquires an extra phase factor which only depends on the evolution path traversed by the system in state space. This phase factor, known as the geometric phase (GP) nowadays \cite{1984Berry45,
1987Aharonov1593,1988Samuel2339,1993Mukunda205}, has been considered a profound and fascinating concept playing a key role in various areas ranging from condensed-matter physics to high energy and particle physics and from quantum information science to gravity and cosmology \cite{2010Xiao1959,2015Sjoeqvist1311,2019Cohen437,2019Bliokh122401,2021Zhao1935}.
In this Letter, we reexamine the GPs accompanying adiabatic evolutions. Unlike Berry who implicitly assumed that the initial state of the system is {perfectly} prepared in one of the eigenstates of the initial Hamiltonian, we take into account small imperfections in the initial-state preparation which are usually, if not always, encountered in practice. We find that the presence of the imperfections can significantly alter the GP in the adiabatic limit $T\rightarrow\infty$, no matter how small the imperfections are. More precisely, if the initial state is perfectly prepared in the eigenstate, the GP converges to the Berry phase in the adiabatic limit, as indicated in Berry's seminal work \cite{1984Berry45} and further confirmed by a number of follow-ups \cite{1988Bender561,1990Wang5107,1995GCS100,2005Tong288}. However, once the initial state slightly deviates from the eigenstate, the GP becomes sensitive to $T$ and does not converge. The above discrepancy leads to the surprising observation that two GPs can differ from each other dramatically even if their evolution paths are sufficiently close in a pointwise manner.
The result of this Letter fills in an important missing piece of the physical picture about the GPs accompanying adiabatic evolutions, which is complementary to but quite different from the piece of the picture discovered by Berry \cite{1984Berry45}.

Let us first recapitulate some fundamentals of the theory of GPs \cite{1984Berry45,1987Aharonov1593,1988Samuel2339,1993Mukunda205}. Consider an $N$-dimensional quantum system. Its evolution can be associated with a path in the state space,
\begin{eqnarray}\label{def-path}
\mathcal{P}: t\in[0,\tau]\rightarrow \ket{\psi(t)}\bra{\psi(t)},
\end{eqnarray}
where $\ket{\psi(t)}$ denotes the evolving state of the system. $\mathcal{P}$
assigns each value of $t \in [0,\tau]$ to the corresponding density operator. Note that Berry originally defined $\mathcal{P}$ to be a curve in the space of parameters on which the Hamiltonian of the system depends \cite{1984Berry45}. It was later realized that $\mathcal{P}$ can be alternatively taken to be a curve in the state space as in Eq.~(\ref{def-path}) \cite{1987Aharonov1593,1988Samuel2339,1993Mukunda205}. This allows for a more compact formulation of the theory of GPs without explicitly referring to the Hamiltonian \cite{1993Mukunda205}.
The GP accompanying $\mathcal{P}$ reads
\begin{eqnarray}\label{def-GP}
\gamma[\mathcal{P}]=\arg\innerproduct{\psi(0)}{\psi(\tau)}+
i
\int_{0}^{\tau}\innerproduct{\psi(t)}
{\partial_t\psi(t)}dt,
\end{eqnarray}
which can be taken as a general definition of the Berry phase applicable to both closed and open paths \cite{1988Samuel2339,1993Mukunda205}.
The geometric nature of definition (\ref{def-GP}) can be seen clearly by inspecting the case of $N=2$. In this case, $\mathcal{P}$ can be pictorially represented by the curve on the Bloch sphere traversed by $\ket{\psi(t)}$. Then, $\gamma[\mathcal{P}]$ equals in magnitude to half the solid angle $\Omega$ subtended by $\mathcal{P}$ at the origin of the Bloch sphere,
\begin{eqnarray}\label{Berry-GP-SA}
\gamma[\mathcal{P}]=-\Omega/2.
\end{eqnarray}
Note that if $\mathcal{P}$ is open, $\Omega$
is determined by the contour that is given by the actual
evolution $\ket{\psi(t)}$ from $\ket{\psi(0)}$ to $\ket{\psi(\tau)}$ and back
along the geodesic curve joining $\ket{\psi(\tau)}$ and $\ket{\psi(0)}$ \cite{1988Samuel2339}.

Keeping the above knowledge in mind, we proceed to the topic of the GP accompanying an adiabatic evolution.
Let $\{H(s), s\in[0,1]\}$ be a given family of Hamiltonians such that $H(0)=H(1)$. Denote the eigenenergies of $H(s)$ by $\varepsilon_0(s),\cdots,\varepsilon_{N-1}(s)$ and the associated eigenstates by $\ket{\varepsilon_0(s)},\cdots,\ket{\varepsilon_{N-1}(s)}$. Here, for simplicity, $\varepsilon_n(s)$'s are assumed to remain distinct for all $s\in[0,1]$, i.e., $\varepsilon_m(s)\neq\varepsilon_n(s)$ whenever $m\neq n$. Besides, without loss of generality, $\ket{\varepsilon_n(s)}$'s are assumed to obey the cyclic condition $\ket{\varepsilon_n(0)}=\ket{\varepsilon_n(1)}$. As usual, by slowly varying the Hamiltonian of the system to run through the given family of Hamiltonians over a long time interval $[0,T]$, we get an adiabatic evolution governed by the Schr\"{o}dinger equation,
$i\partial_t\ket{\psi(t)}=H(s(t))\ket{\psi(t)}$,
with $s(t)=t/T$. Note that $\ket{\psi(t)}$ depends on $T$ which controls the evolution rate. For convenience, we rewrite the Schr\"{o}dinger equation as
\begin{eqnarray}\label{SE-s}
i\partial_s\ket{\psi_T(s)}=TH(s)\ket{\psi_T(s)}.
\end{eqnarray}
Here, the subscript $T$ has been added, to distinguish $\ket{\psi_T(s)}$ from $\ket{\psi(t)}$. The evolution operator associated with Eq.~(\ref{SE-s}) is denoted by $U_T(s)$.

We first examine the setting that
the initial state of the system is perfectly prepared in one of the eigenstates of $H(0)$, say, $\ket{\varepsilon_0(0)}$. That is,
\begin{eqnarray}\label{initial-state-perfect}
\ket{\psi_T(0)}=\ket{\varepsilon_0(0)}.
\end{eqnarray}
Evidently, this setting is essentially the same as that considered by Berry \cite{1984Berry45}, which is referred to as the perfect setting hereafter. Assuming that $T$ is sufficiently large, we can neglect non-adiabatic effects and express $\ket{\psi_T(s)}$ as $\exp[i\alpha (s)]\ket{\varepsilon_0(s)}$. Inserting this expression into Eq.~(\ref{SE-s}) and following the same arguments in Ref.~\cite{1984Berry45}, we have
\begin{eqnarray}\label{evolving-state-perfect}
\ket{\psi_T(s)}=e^{-iT\int_0^s\varepsilon_0(\sigma)d\sigma}
e^{i\gamma_0(s)}\ket{\varepsilon_0(s)}.
\end{eqnarray}
Here, $\gamma_n(s)=i\int_{0}^{s}\innerproduct{\varepsilon_n(\sigma)}
{\partial_\sigma\varepsilon_n(\sigma)}d\sigma$. Equation (\ref{evolving-state-perfect}) defines the evolution path
\begin{eqnarray}
\mathcal{P}_T: s\in[0,1]\rightarrow \ket{\psi_T(s)}\bra{\psi_T(s)}.
\end{eqnarray}
Here, the subscript $T$ has been added in $\mathcal{P}_T$, though $\mathcal{P}_T$ does not depend on $T$. Note that if non-adiabatic effects are taken into account, $\mathcal{P}_T$ would slightly depend on $T$ \cite{1988Bender561,1990Wang5107,1995GCS100,2005Tong288}.
Substituting Eq.~(\ref{evolving-state-perfect}) into Eq.~(\ref{def-GP}) yields the GP in the perfect setting,
\begin{eqnarray}\label{GP-perfect}
\gamma[\mathcal{P}_T]=\gamma_0(1).
\end{eqnarray}
Clearly, $\gamma[\mathcal{P}_T]$ is nothing but the Berry phase associated with $\ket{\varepsilon_0(s)}$.

We then examine the setting that there exist small imperfections in the initial-state preparation. That is,
\begin{eqnarray}\label{initial-state-imperfect}
\ket{\psi_T^\prime(0)}=\sum_{n=0}^{N-1}a_n\ket{\varepsilon_n(0)},
\end{eqnarray}
with
\begin{eqnarray}
\abs{a_0}\approx 1~~ \textrm{and}~~ \abs{a_n}\ll 1 ~~\textrm{for}~~ n\neq 0.
\end{eqnarray}
Here and henceforth, primes are placed on notations like $\ket{\psi_T^\prime(0)}$, in order to distinguish the above setting, which is referred to as the imperfect setting, from the perfect setting. Note that $U_T(s)$ maps the state $\ket{\varepsilon_n(0)}$ into the state $e^{-iT\int_0^s\varepsilon_n(\sigma)d\sigma}
e^{i\gamma_n(s)}\ket{\varepsilon_n(s)}$. Linearity of $U_T(s)$ implies that the evolving state in the imperfect setting reads
\begin{eqnarray}\label{evolving-state-imperfect}
\ket{\psi_T^\prime(s)}=\sum_{n=0}^{N-1}a_ne^{-iT\int_0^s\varepsilon_n(\sigma)d\sigma}
e^{i\gamma_n(s)}\ket{\varepsilon_n(s)}.
\end{eqnarray}
Equation (\ref{evolving-state-imperfect}) defines the evolution path
\begin{eqnarray}
\mathcal{P}_T^\prime: s\in[0,1]\rightarrow \ket{\psi_T^\prime(s)}\bra{\psi_T^\prime(s)}.
\end{eqnarray}
Using $\ket{\psi_T(s)}=U_T(s)\ket{\psi_T(0)}$ and $\ket{\psi_T^\prime(s)}=U_T(s)\ket{\psi_T^\prime(0)}$, we have
\begin{eqnarray}\label{eq:colseness}
\abs{\innerproduct{\psi_T(s)}{\psi_T^\prime(s)}}
=\abs{\innerproduct{\psi_T(0)}{\psi_T^\prime(0)}}=\abs{a_0}
\approx 1,
\end{eqnarray}
which holds for any value of $T$.  Equation (\ref{eq:colseness}) implies that $\mathcal{P}_T$ and $\mathcal{P}_T^\prime$ are close to each other in a pointwise manner. Evidently, the smaller the imperfections $\abs{a_n}$'s, $n\neq 0$, are, the closer the two paths become. Notably, a widely-held intuition is that two GPs should be approximately equally large when their evolution paths are pointwisely close \cite{2003Chiara90404,2003Carollo160402,2004Solinas42316,2005Zhu20301,
2005FuentesGuridi20503,2006Florio22327,2011Thomas42335}. Accordingly, the relation $\gamma[\mathcal{P}_T]\approx\gamma[\mathcal{P}_T^\prime]$ shall hold when the imperfections are small enough. At first glance, the relation seems to be consistent with the geometric picture established by Berry \cite{1984Berry45}, which ties the values of GPs to the solid angles subtended by their evolution paths as in Eq.~(\ref{Berry-GP-SA}). Moreover, the relation seems also reasonable from a physical point of view, since Eq.~(\ref{eq:colseness}) means that the two states $\ket{\psi_T(t)}\bra{\psi_T(t)}$ and $\ket{\psi_T^\prime(t)}\bra{\psi_T^\prime(t)}$ are hardly distinguishable according to quantum mechanics during the whole evolution. However, we find that the relation does not hold in general, no matter how small the imperfections are.
Inserting Eq.~(\ref{evolving-state-imperfect}) into Eq.~(\ref{def-GP}), we reach the key formula \cite{SM}:
\begin{eqnarray}\label{GP-imperfect}
\gamma[\mathcal{P}_T^\prime]\approx\gamma_0(1)+\sum_{n\neq 0}\abs{a_n}^2T
\int_0^1\left[\varepsilon_n(s)-\varepsilon_0(s)\right]ds,
\end{eqnarray}
stating that the GP in the imperfect setting approximately equals to the Berry phase plus a correction term up to an integer multiple of $2\pi$.

Now, with Eqs.~(\ref{GP-perfect}) and (\ref{GP-imperfect}), we are able to compare $\gamma[\mathcal{P}_T]$ and $\gamma[\mathcal{P}_T^\prime]$. Apparently, $\gamma[\mathcal{P}_T]$ is solely determined by the eigenprojection $\ket{\varepsilon_0(s)}\bra{\varepsilon_0(s)}$, irrespective of the value of $T$ as well as the eigenenergies $\varepsilon_n(s)$. This is just the conventional wisdom that GPs are geometric in nature and is insensitive to evolution details. In sharp contrast, $\gamma[\mathcal{P}_T^\prime]$ is sensitive to $T$, due to the emergence of the correction term. Indeed, no matter how small the imperfections are, the correction term can take any non-negligible value with a proper choice of $T$. Besides, it is easy to see that $\gamma[\mathcal{P}_T^\prime]$ is sensitive to $\varepsilon_n(s)$ when $T$ is large. The above comparison uncovers the striking result that, whereas the GP is insensitive to $T$ as well as $\varepsilon_n(s)$ in the perfect setting as indicated in many previous works \cite{1984Berry45,1988Bender561,1990Wang5107,1995GCS100,2005Tong288}, the GP becomes sensitive to $T$ as well as $\varepsilon_n(s)$ in the imperfect setting.
As such, it is possible for $\gamma[\mathcal{P}_T]$  and $\gamma[\mathcal{P}_T^\prime]$ to  differ from each other dramatically even if $\mathcal{P}_T$ and $\mathcal{P}_T^\prime$ are sufficiently close in a pointwise manner.
We emphasize that the anomalous behavior of the GP found here is not a consequence of the inconsistency in the application of the adiabatic theorem first reported by Marzlin and Sanders \cite{2004Marzlin160408}.
As a matter of fact, the physical context under consideration, i.e., adiabatic evolutions with imperfect initial-state preparations, is not pathological from both theoretical and practical points of view.

Let us furnish an exactly solvable model to substantiate the result obtained above. Consider the well-known model of a spin-half particle in a rotating magnetic field. Its Hamiltonian reads
\begin{eqnarray}
H(s)=&&-\frac{\omega_0}{2}\left[\sigma_x\sin\theta\cos(2\pi s)+\sigma_y\sin\theta\sin(2\pi s)
\right.\nonumber\\
&&\left.+\sigma_z\cos\theta\right],
\end{eqnarray}
where $\omega_0$ is a real parameter defined by the magnetic moment of the spin and the intensity of the external magnetic field, and $\sigma_i$, $i=x,y,z$, denote the Pauli matrices. The instantaneous eigenvalues and eigenstates of $H(s)$ are given by
\begin{eqnarray}\label{example-eigensystem}
\varepsilon_0(s)=-\frac{\omega_0}{2},~~~~~~
\ket{\varepsilon_0(s)}=
\begin{pmatrix}
  \cos\frac{\theta}{2} \\
  \sin\frac{\theta}{2}e^{i2\pi s}
\end{pmatrix};\nonumber\\
\varepsilon_1(s)=\frac{\omega_0}{2},~~~~~~
\ket{\varepsilon_1(s)}=
\begin{pmatrix}
  \sin\frac{\theta}{2} \\
  -\cos\frac{\theta}{2}e^{i2\pi s}
\end{pmatrix}.
\end{eqnarray}
Solving the Schr\"{o}dinger equation (\ref{SE-s}), we can find the evolution operator
\begin{eqnarray}
U_T(s)=e^{-i\pi s\sigma_z}e^{i\frac{\overline{\omega}Ts}{2}\left(
\frac{\omega_0\sin\theta}{\overline{\omega}}\sigma_x+
\frac{\omega_0\cos\theta+\omega}{\overline{\omega}}
\sigma_z\right)},
\end{eqnarray}
where $\omega=2\pi/T$ and
$\overline{\omega}=\sqrt{\omega_0^2+2\omega_0\omega\cos\theta+\omega^2}$.
Applying $U_T(s)$ to $\ket{\varepsilon_0(0)}$, we have the evolving state in the perfect setting
\begin{eqnarray}\label{example-evolving-state-perfect}
\ket{\psi_T(s)}=
\begin{pmatrix}
\left(\cos\frac{\overline{\omega}Ts}{2}+i\frac{\omega_0+\omega}{\overline{\omega}}
\sin\frac{\overline{\omega}Ts}{2}\right)\cos\frac{\theta}{2}e^{-i\pi s} \\
\left(\cos\frac{\overline{\omega}Ts}{2}+i\frac{\omega_0-\omega}{\overline{\omega}}
\sin\frac{\overline{\omega}Ts}{2}\right)\sin\frac{\theta}{2}e^{i\pi s}
\end{pmatrix}.\nonumber\\
\end{eqnarray}
Substituting Eq.~(\ref{example-evolving-state-perfect}) into Eq.~(\ref{def-GP}), we obtain
\begin{eqnarray}\label{example-GP-perfect}
&&\gamma[\mathcal{P}_T]=\arg\left(-\cos\frac{\overline{\omega}T}{2}-
i\frac{\omega_0+\omega\cos\theta}{\overline{\omega}}\sin\frac{\overline{\omega}T}{2}\right)
-\nonumber\\
&&\frac{\omega_0 T}{2}\left[1-\frac{\omega^2{\sin^2\theta}}{\overline{\omega}^2}
\left(1-\frac{\sin(\overline{\omega}T)}{\overline{\omega}T}\right)\right].
\end{eqnarray}
Likewise, applying $U_T(s)$ to $a_0\ket{\varepsilon_0(0)}+a_1\ket{\varepsilon_1(0)}$, we have the evolving state in the imperfect setting
\begin{eqnarray}\label{example-evolving-state-imperfect}
&&\ket{\psi_T^\prime(s)}=
a_0
\begin{pmatrix}
\left(\cos\frac{\overline{\omega}Ts}{2}+i\frac{\omega_0+\omega}{\overline{\omega}}
\sin\frac{\overline{\omega}Ts}{2}\right)\cos\frac{\theta}{2}e^{-i\pi s} \\
\left(\cos\frac{\overline{\omega}Ts}{2}+i\frac{\omega_0-\omega}{\overline{\omega}}
\sin\frac{\overline{\omega}Ts}{2}\right)\sin\frac{\theta}{2}e^{i\pi s}
\end{pmatrix}
+\nonumber\\
&&a_1
\begin{pmatrix}
\left(\cos\frac{\overline{\omega}Ts}{2}-i\frac{\omega_0-\omega}{\overline{\omega}}
\sin\frac{\overline{\omega}Ts}{2}\right)\sin\frac{\theta}{2}e^{-i\pi s} \\
-\left(\cos\frac{\overline{\omega}Ts}{2}-i\frac{\omega_0+\omega}{\overline{\omega}}
\sin\frac{\overline{\omega}Ts}{2}\right)\cos\frac{\theta}{2}e^{i\pi s}
\end{pmatrix}.
\end{eqnarray}
Inserting Eq.~(\ref{example-evolving-state-imperfect}) into Eq.~(\ref{def-GP}), we obtain
\begin{eqnarray}\label{example-GP-imperfect}
&&\gamma[\mathcal{P}_T^\prime]=
\arg\left(-\cos\frac{\overline{\omega}T}{2}
-i\frac{2\Re(a_0^*a_1)\omega\sin\theta}{\overline{\omega}}
\sin\frac{\overline{\omega}T}{2}-\right.\nonumber\\
&&\left.i\frac{(\abs{a_0}^2-\abs{a_1}^2)
(\omega_0+\omega\cos\theta)}{\overline{\omega}}
\sin\frac{\overline{\omega}T}{2}\right)-\frac{\omega_0T}{2}(\abs{a_0}^2-\nonumber\\
&&\abs{a_1}^2)\left[1-\frac{\omega^2{\sin^2\theta}}{\overline{\omega}^2}
\left(1-\frac{\sin(\overline{\omega}T)}{\overline{\omega}T}\right)\right]-
\frac{2\pi\omega_0\sin\theta}{\overline{\omega}}\times\nonumber\\
&&\left[\Re(a_0^*a_1)\frac{\omega_0+\omega\cos\theta}{\overline{\omega}}
\left(1-\frac{\sin(\overline{\omega}T)}{\overline{\omega}T}\right)
-\Im(a_0^*a_1)\times\right.\nonumber\\
&&\left.\frac{1-\cos(\overline{\omega}T)}{\overline{\omega}T}\right].
\end{eqnarray}
It is worth noting that Eqs.~(\ref{example-GP-perfect}) and (\ref{example-GP-imperfect}) are exact results without any approximation. In particular, the non-adiabatic effects have been taken into account in these two equations.

\begin{figure}
\includegraphics[width=0.4\textwidth]{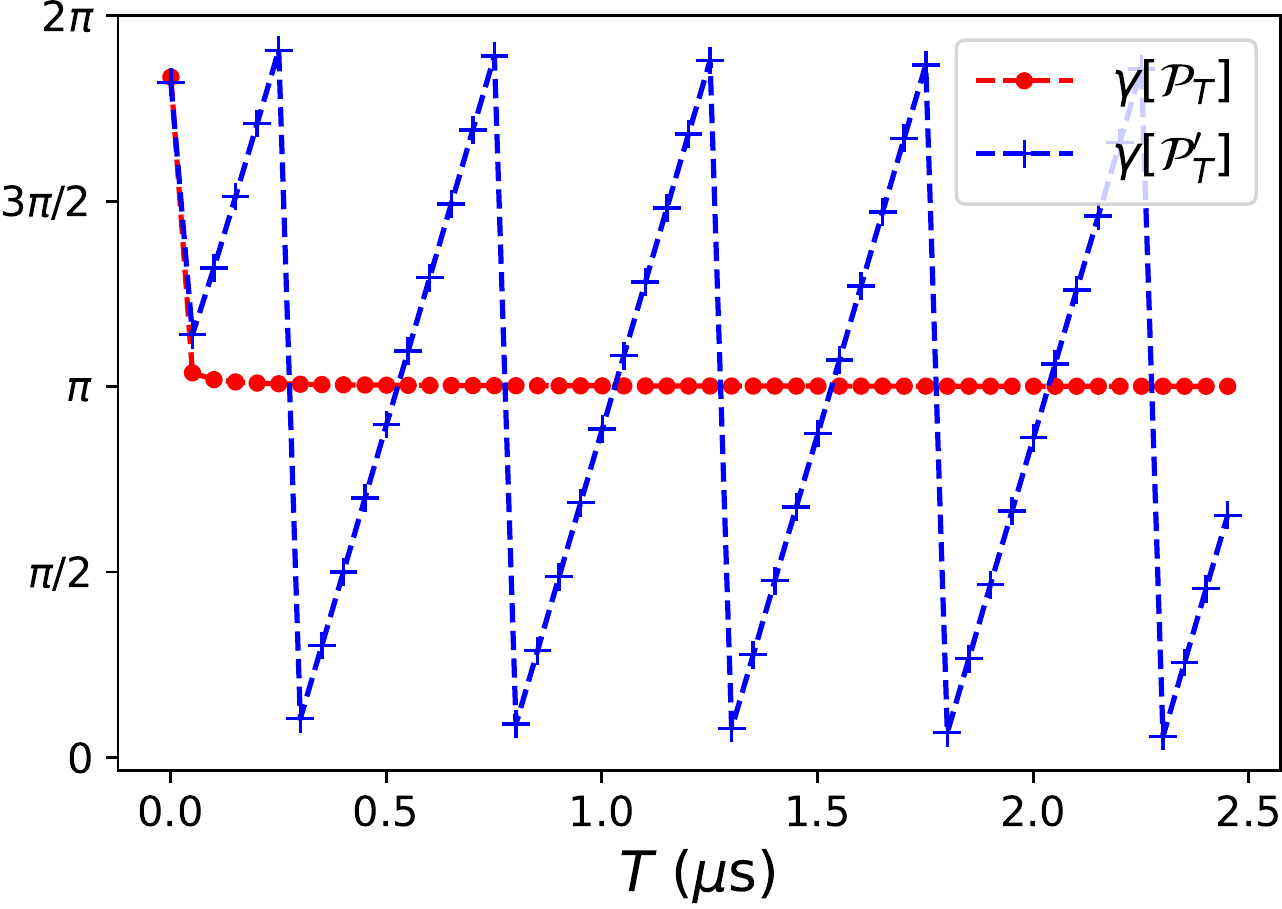}
\caption{Numerical results of the geometric phases given in Eqs.~(\ref{example-GP-perfect}) and (\ref{example-GP-imperfect}) as functions of $T$. The red dots and blue crosses correspond to $\gamma[\mathcal{P}_T]$ and $\gamma[\mathcal{P}_T^\prime]$, respectively. Parameters used are $\theta=\pi/2$, $\omega_0=5$ GHz, $a_0=\sqrt{399/400}$, and $a_1=\sqrt{1/400}$.}
\label{fig1}
\end{figure}

To clearly see the difference between $\gamma[\mathcal{P}_T]$ and $\gamma[\mathcal{P}_T^\prime]$, we neglect the small terms in Eqs.~(\ref{example-GP-perfect}) and (\ref{example-GP-imperfect}) such as the terms of orders equal to or higher than $\mathcal{O}(1/T)$ , and obtain
\begin{eqnarray}\label{example-GP-perfect-app}
\gamma[\mathcal{P}_T]\approx -\pi(1-\cos\theta),
\end{eqnarray}
and
\begin{eqnarray}\label{example-GP-imperfect-app}
\gamma[\mathcal{P}_T^\prime]\approx-\pi(1-\cos\theta)+\abs{a_1}^2T\omega_0.
\end{eqnarray}
Evidently, Eqs.~(\ref{example-GP-perfect-app}) and (\ref{example-GP-imperfect-app}) are in accordance with Eqs.~(\ref{GP-perfect}) and (\ref{GP-imperfect}), respectively. Noting that $-\pi(1-\cos\theta)$ is the Berry phase associated with $\ket{\varepsilon_0(s)}$ appearing in Eq.~(\ref{example-eigensystem}), we deduce from Eq.~(\ref{example-GP-perfect-app}) that $\gamma[\mathcal{P}_T]$ approaches the Berry phase as $T\rightarrow\infty$ and is insensitive to $T$ when $T$ is sufficiently large. In contrast, $\gamma[\mathcal{P}_T^\prime]$ is sensitive to $T$ since its value continuously varies in the course of increasing $T$, as can be seen from Eq.~(\ref{example-GP-imperfect-app}). To further confirm these analytical results, we numerically figure out $\gamma[\mathcal{P}_T]$ and $\gamma[\mathcal{P}_T^\prime]$ for different $T$ using the exact equations (\ref{example-GP-perfect}) and (\ref{example-GP-imperfect}).
The numerical results are presented in Fig.~\ref{fig1}. Here, in view of the experimental studies \cite{2000Jones869,2001Duan1695,2014Zu72}, we set $\omega_0=5$ GHz. Besides, we choose $a_0=\sqrt{399/400}$, $a_1=\sqrt{1/400}$, and $\theta=\pi/2$. Considering that a phase factor is defined up to an integer multiple of $2\pi$, we set the plot range of the vertical axis to be $[0,2\pi]$.
Clearly, as $T$ increases, $\gamma[\mathcal{P}_T]$ converges to $\pi$ (see the red dots), whereas $\gamma[\mathcal{P}_T^\prime]$ goes up and down and approaches all the values in $[0,2\pi]$ (see the blue crosses). Note that $\abs{\innerproduct{\psi_T(s)}{\psi_T^\prime(s)}}
=\abs{a_0}=99.87\%$, implying that $\mathcal{P}_T$ and $\mathcal{P}_T^\prime$ are pointwisely close as two curves on the Bloch sphere. Therefore, the numerical results, which are consistent with the above analytical ones, demonstrate that $\gamma[\mathcal{P}_T]$ and $\gamma[\mathcal{P}_T^\prime]$ can differ from each other dramatically even if $\mathcal{P}_T$ and $\mathcal{P}_T^\prime$ are very close in a pointwise manner.

\begin{figure}
\includegraphics[width=0.4\textwidth]{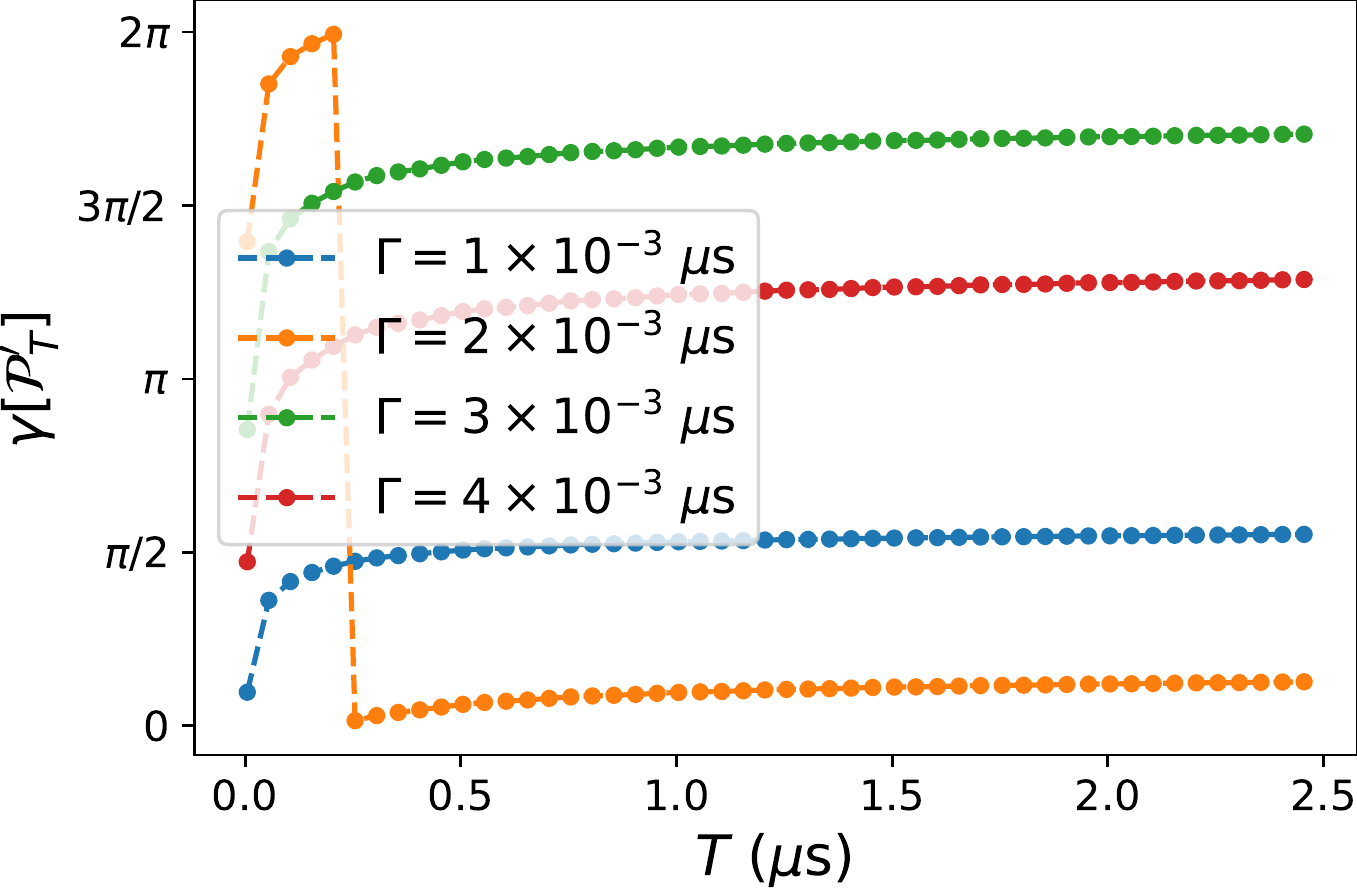}
\caption{Numerical results of $\gamma[\mathcal{P}_T^\prime]$ in Eq.~(\ref{example-GP-imperfect}) as a function of $T$ for different $\Gamma$. Parameters used are $\theta=\pi/2$, $\omega_0=5$ GHz, $a_0=\sqrt{1-\Gamma/T}$, and $a_1=\sqrt{\Gamma/T}$, with $\Gamma$ given in the figure legend.}
\label{fig2}
\end{figure}

Up to now, we have implicitly assumed that $\abs{a_n}$'s are independent of $T$. It is interesting to inspect the case that $\abs{a_n}$'s are dependent of $T$. The $T$-dependence of $\abs{a_n}$'s may arise in the situation that $\ket{\psi_T^\prime(0)}$ is prepared via some adiabatic passage executed before $t=0$ \cite{2001Farhi472,2014Zhang42321}, whose evolution rate is chosen to be also related to $T$. Let
${a_0}=\sqrt{1-\Gamma/T}$ and ${a_1}=\sqrt{\Gamma/T}$,
where $\Gamma$ is a fixed positive number. Then, ${a_0}\rightarrow 1$ and ${a_1}\rightarrow 0$ as $T\rightarrow\infty$, implying that $\lim_{T\rightarrow\infty}\ket{\psi_T^\prime(s)}\bra{\psi_T^\prime(s)}=
\ket{\varepsilon_0(s)}\bra{\varepsilon_0(s)}$ for all $s\in[0,1]$. Besides, $\lim_{T\rightarrow\infty}\ket{\psi_T(s)}\bra{\psi_T(s)}=
\ket{\varepsilon_0(s)}\bra{\varepsilon_0(s)}$ for all $s\in[0,1]$. Thus, in the limit of $T\rightarrow\infty$, $\mathcal{P}_T$ and $\mathcal{P}_T^\prime$ converge to the same adiabatic path $\mathcal{P}_\textrm{adi}:s\in[0,1]\rightarrow\ket{\varepsilon_0(s)}\bra{\varepsilon_0(s)}$,
\begin{eqnarray}\label{ex:PP}
\lim_{T\rightarrow\infty}\mathcal{P}_T=\mathcal{P}_\textrm{adi}
=\lim_{T\rightarrow\infty}\mathcal{P}_T^\prime.
\end{eqnarray}
It follows from Eqs.~(\ref{example-GP-perfect-app}) and (\ref{example-GP-imperfect-app}) that $\lim_{T\rightarrow\infty}\gamma[\mathcal{P}_T]=-\pi(1-\cos\theta)$ and $\lim_{T\rightarrow\infty}\gamma[\mathcal{P}_T^\prime]=-\pi(1-\cos\theta)+\Gamma\omega_0$. Noting that $\Gamma\omega_0$ can take any value with a proper choice of $\Gamma$, we have
\begin{eqnarray}\label{ex:GG}
\lim_{T\rightarrow\infty}\gamma[\mathcal{P}_T]=\gamma_0(1)\neq \lim_{T\rightarrow\infty}\gamma[\mathcal{P}_T^\prime].
\end{eqnarray}
It follows from Eqs.~(\ref{ex:PP}) and (\ref{ex:GG}) that, in the limit of $T\rightarrow\infty$, $\mathcal{P}_T$ and $\mathcal{P}_T^\prime$ are \textit{arbitrarily} close to each other, but $\gamma[\mathcal{P}_T]$ and $\gamma[\mathcal{P}_T^\prime]$ can still distinguish from each other dramatically. Figure \ref{fig2} shows the numerical results of $\gamma[\mathcal{P}_T^\prime]$ for different $\Gamma$, which are computed from the exact equation (\ref{example-GP-imperfect}). Notably, as $T\rightarrow\infty$, $\mathcal{P}_T^\prime$ always converges to $\mathcal{P}_\textrm{adi}$ regardless of the value of $\Gamma$, but $\gamma[\mathcal{P}_T^\prime]$ converges to a different value for a different $\Gamma$, as can be seen from Fig.~\ref{fig2}. This indicates that the numerical results are consistent with the above analytical ones.

\begin{figure}
\includegraphics[width=0.40\textwidth]{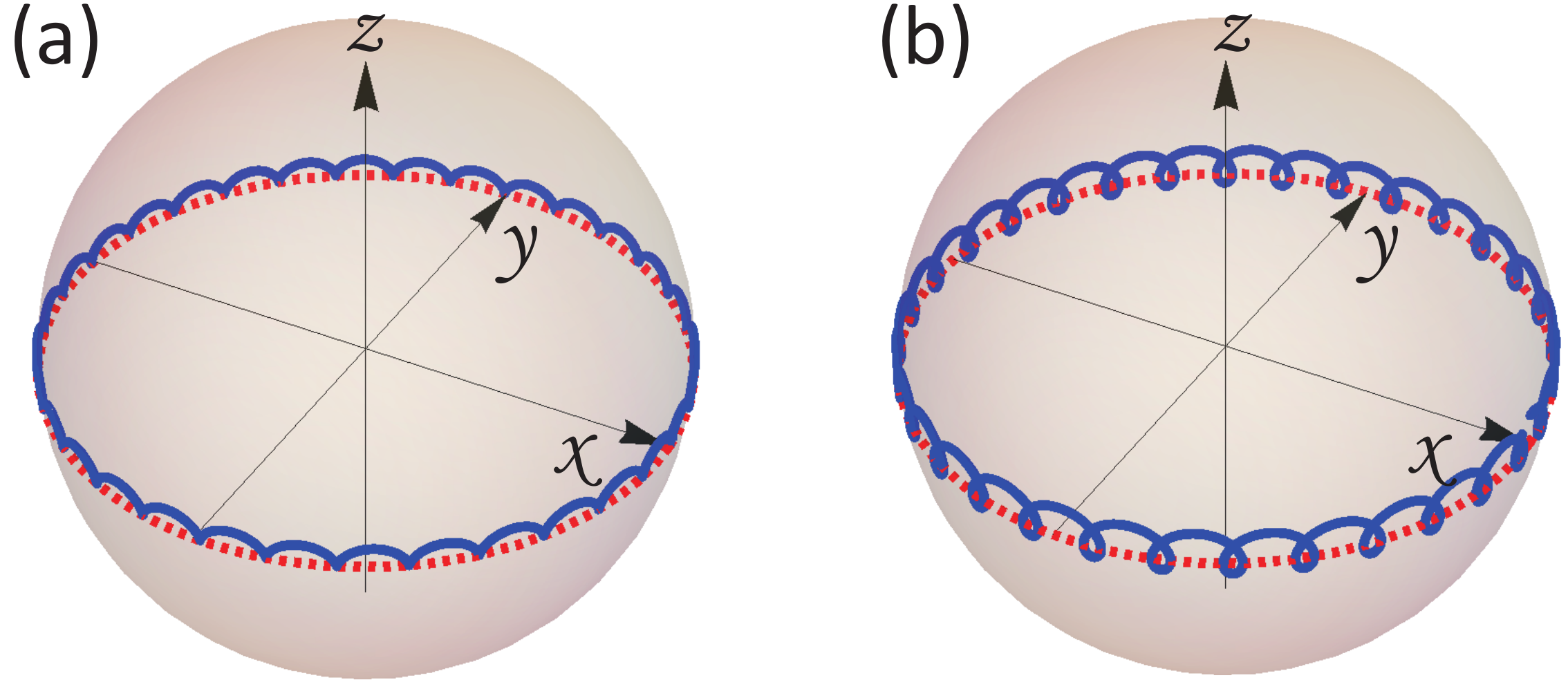}
\caption{Illustration of (a) $\mathcal{P}_T$ and (b) $\mathcal{P}_T^\prime$ as two curves on the Bloch sphere. The red dashed circles in both subfigures correspond to the adiabatic path $\mathcal{P}_\textrm{adi}: s\in[0,1]\rightarrow\ket{\varepsilon_0(s)}\bra{\varepsilon_0(s)}$ with $\ket{\varepsilon_0(s)}$ given in Eq.~(\ref{example-eigensystem}). Parameters used are $\theta=\pi/2$, $\omega_0=5$ GHz, $T=0.04$ $\mu$s, $a_0=\sqrt{399/400}$, and $a_1=\sqrt{1/400}$.}
\label{fig3}
\end{figure}

Lastly, it is instructive to give a geometric picture for comprehending the result that $\gamma[\mathcal{P}_T]$ and $\gamma[\mathcal{P}_T^\prime]$ can differ dramatically from each other, no matter how small $\abs{a_1}$ is. $\mathcal{P}_T$ is a simple curve on the Bloch sphere (see the blue solid curve in Fig.~\ref{fig3}a), for which Stokes' theorem holds and we have
\begin{eqnarray}
\gamma[\mathcal{P}_T]=-\Omega/2,
\end{eqnarray}
for a sufficiently large $T$, where $\Omega$ is the solid angle subtended by $\mathcal{P}_T$ at the origin of the Bloch sphere. This point can be verified by inspecting Eq.~(\ref{example-GP-perfect-app}) and noting that $\Omega=2\pi(1-\cos\theta)$. However, unlike $\mathcal{P}_T$, $\mathcal{P}_T^\prime$ is not simple and crosses itself many times (see the blue solid curve in Fig.~\ref{fig3}b). That is, the presence of the imperfections changes the nature of the evolution path from being a simple curve to being a self-crossing curve. The self-crossing leads to many small circles. The number of the circles is $\omega_0T/(2\pi)$ and the signed surface area enclosed by each circle is $-4\pi\abs{a_1}^2$. So, due to the appearance of these circles, the solid angle associated with $\mathcal{P}_T^\prime$ should be corrected as $\Omega^\prime=\Omega-4\pi\abs{a_1}^2\times\omega_0T/(2\pi)=\Omega-2\abs{a_1}^2\omega_0T$. Then, we have
\begin{eqnarray}
\gamma[\mathcal{P}_T^\prime]=-\Omega^\prime/2,
\end{eqnarray}
for a sufficiently large $T$. Now, it is clear that the self-crossing exhibited by $\mathcal{P}_T^\prime$ results in the correction term appearing in Eq.~(\ref{example-GP-imperfect-app}).

In summary, we have found that, whereas the GP is insensitive to $T$ in the perfect setting when $T$ is large as shown in many previous works, it becomes sensitive to $T$ once the initial state deviates from the eigenstate. We have shown that such an anomalous behavior is due to the emergence of the correction term, which stems from the self-crossings of the evolution path and can take any non-negligible value no matter how small the imperfections are. Astonishingly, this means that even an infinitesimal imperfection can yield a significant change of the GP in the adiabatic limit. As such, contrary to the widely-held intuition, two GPs can dramatically differ from each other even if their evolution paths are sufficiently close in a pointwise manner, which represents an important missing piece of the physical picture about the GPs accompanying adiabatic evolutions. The result obtained here is a necessary complement to Berry's remarkable discovery.

\textit{Note added.}---Very recently, Zhu, Lu, and Lein found that contrary to expectations, the Berry phase and the Aharonov-Anandan phase may not coincide in the adiabatic limit when their evolution paths lead through degeneracies of energy levels \cite{Zhu}.
Interestingly enough, our result reveals that GPs do not behave as expected even in the familiar and pervasive context of adiabatic evolutions without degeneracies.

\begin{acknowledgments}
This work was supported by the National Natural Science Foundation of
China through Grant Nos.~11705105 and 11775129.
\end{acknowledgments}


\begin{thebibliography}{28}%
\makeatletter
\providecommand \@ifxundefined [1]{%
 \@ifx{#1\undefined}
}%
\providecommand \@ifnum [1]{%
 \ifnum #1\expandafter \@firstoftwo
 \else \expandafter \@secondoftwo
 \fi
}%
\providecommand \@ifx [1]{%
 \ifx #1\expandafter \@firstoftwo
 \else \expandafter \@secondoftwo
 \fi
}%
\providecommand \natexlab [1]{#1}%
\providecommand \enquote  [1]{``#1''}%
\providecommand \bibnamefont  [1]{#1}%
\providecommand \bibfnamefont [1]{#1}%
\providecommand \citenamefont [1]{#1}%
\providecommand \href@noop [0]{\@secondoftwo}%
\providecommand \href [0]{\begingroup \@sanitize@url \@href}%
\providecommand \@href[1]{\@@startlink{#1}\@@href}%
\providecommand \@@href[1]{\endgroup#1\@@endlink}%
\providecommand \@sanitize@url [0]{\catcode `\\12\catcode `\$12\catcode
  `\&12\catcode `\#12\catcode `\^12\catcode `\_12\catcode `\%12\relax}%
\providecommand \@@startlink[1]{}%
\providecommand \@@endlink[0]{}%
\providecommand \url  [0]{\begingroup\@sanitize@url \@url }%
\providecommand \@url [1]{\endgroup\@href {#1}{\urlprefix }}%
\providecommand \urlprefix  [0]{URL }%
\providecommand \Eprint [0]{\href }%
\providecommand \doibase [0]{http://dx.doi.org/}%
\providecommand \selectlanguage [0]{\@gobble}%
\providecommand \bibinfo  [0]{\@secondoftwo}%
\providecommand \bibfield  [0]{\@secondoftwo}%
\providecommand \translation [1]{[#1]}%
\providecommand \BibitemOpen [0]{}%
\providecommand \bibitemStop [0]{}%
\providecommand \bibitemNoStop [0]{.\EOS\space}%
\providecommand \EOS [0]{\spacefactor3000\relax}%
\providecommand \BibitemShut  [1]{\csname bibitem#1\endcsname}%
\let\auto@bib@innerbib\@empty
\bibitem [{\citenamefont {Berry}(1984)}]{1984Berry45}%
  \BibitemOpen
  \bibfield  {author} {\bibinfo {author} {\bibfnamefont {M.~V.}\ \bibnamefont
  {Berry}},\ }\bibfield  {title} {\enquote {\bibinfo {title} {Quantal phase
  factors accompanying adiabatic changes},}\ }\href {\doibase
  10.1142/9789813221215_0006} {\bibfield  {journal} {\bibinfo  {journal} {Proc.
  R. Soc. Lond. A}\ }\textbf {\bibinfo {volume} {392}},\ \bibinfo {pages} {45}
  (\bibinfo {year} {1984})}\BibitemShut {NoStop}%
\bibitem [{\citenamefont {Aharonov}\ and\ \citenamefont
  {Anandan}(1987)}]{1987Aharonov1593}%
  \BibitemOpen
  \bibfield  {author} {\bibinfo {author} {\bibfnamefont {Y.}~\bibnamefont
  {Aharonov}}\ and\ \bibinfo {author} {\bibfnamefont {J.}~\bibnamefont
  {Anandan}},\ }\bibfield  {title} {\enquote {\bibinfo {title} {Phase change
  during a cyclic quantum evolution},}\ }\href {\doibase
  10.1103/PhysRevLett.58.1593} {\bibfield  {journal} {\bibinfo  {journal}
  {Phys. Rev. Lett.}\ }\textbf {\bibinfo {volume} {58}},\ \bibinfo {pages}
  {1593} (\bibinfo {year} {1987})}\BibitemShut {NoStop}%
\bibitem [{\citenamefont {Samuel}\ and\ \citenamefont
  {Bhandari}(1988)}]{1988Samuel2339}%
  \BibitemOpen
  \bibfield  {author} {\bibinfo {author} {\bibfnamefont {J.}~\bibnamefont
  {Samuel}}\ and\ \bibinfo {author} {\bibfnamefont {R.}~\bibnamefont
  {Bhandari}},\ }\bibfield  {title} {\enquote {\bibinfo {title} {General
  setting for {Berry's} phase},}\ }\href {\doibase 10.1103/PhysRevLett.60.2339}
  {\bibfield  {journal} {\bibinfo  {journal} {Phys. Rev. Lett.}\ }\textbf
  {\bibinfo {volume} {60}},\ \bibinfo {pages} {2339} (\bibinfo {year}
  {1988})}\BibitemShut {NoStop}%
\bibitem [{\citenamefont {Mukunda}\ and\ \citenamefont
  {Simon}(1993)}]{1993Mukunda205}%
  \BibitemOpen
  \bibfield  {author} {\bibinfo {author} {\bibfnamefont {N.}~\bibnamefont
  {Mukunda}}\ and\ \bibinfo {author} {\bibfnamefont {R.}~\bibnamefont
  {Simon}},\ }\bibfield  {title} {\enquote {\bibinfo {title} {Quantum kinematic
  approach to the geometric phase. i. general formalism},}\ }\href
  {https://doi.org/10.1006/aphy.1993.1093} {\bibfield  {journal} {\bibinfo
  {journal} {Ann. Phys. (N.Y.)}\ }\textbf {\bibinfo {volume} {228}},\ \bibinfo
  {pages} {205} (\bibinfo {year} {1993})}\BibitemShut {NoStop}%
\bibitem [{\citenamefont {Xiao}\ \emph {et~al.}(2010)\citenamefont {Xiao},
  \citenamefont {Chang},\ and\ \citenamefont {Niu}}]{2010Xiao1959}%
  \BibitemOpen
  \bibfield  {author} {\bibinfo {author} {\bibfnamefont {D.}~\bibnamefont
  {Xiao}}, \bibinfo {author} {\bibfnamefont {M.-C.}\ \bibnamefont {Chang}}, \
  and\ \bibinfo {author} {\bibfnamefont {Q.}~\bibnamefont {Niu}},\ }\bibfield
  {title} {\enquote {\bibinfo {title} {Berry phase effects on electronic
  properties},}\ }\href {\doibase 10.1103/RevModPhys.82.1959} {\bibfield
  {journal} {\bibinfo  {journal} {Rev. Mod. Phys.}\ }\textbf {\bibinfo {volume}
  {82}},\ \bibinfo {pages} {1959} (\bibinfo {year} {2010})}\BibitemShut
  {NoStop}%
\bibitem [{\citenamefont {Sjöqvist}(2015)}]{2015Sjoeqvist1311}%
  \BibitemOpen
  \bibfield  {author} {\bibinfo {author} {\bibfnamefont {E.}~\bibnamefont
  {Sjöqvist}},\ }\bibfield  {title} {\enquote {\bibinfo {title} {Geometric
  phases in quantum information},}\ }\href {\doibase 10.1002/qua.24941}
  {\bibfield  {journal} {\bibinfo  {journal} {Int. J. Quant. Chem.}\ }\textbf
  {\bibinfo {volume} {115}},\ \bibinfo {pages} {1311} (\bibinfo {year}
  {2015})}\BibitemShut {NoStop}%
\bibitem [{\citenamefont {Cohen}\ \emph {et~al.}(2019)\citenamefont {Cohen},
  \citenamefont {Larocque}, \citenamefont {Bouchard}, \citenamefont
  {Nejadsattari}, \citenamefont {Gefen},\ and\ \citenamefont
  {Karimi}}]{2019Cohen437}%
  \BibitemOpen
  \bibfield  {author} {\bibinfo {author} {\bibfnamefont {E.}~\bibnamefont
  {Cohen}}, \bibinfo {author} {\bibfnamefont {H.}~\bibnamefont {Larocque}},
  \bibinfo {author} {\bibfnamefont {F.}~\bibnamefont {Bouchard}}, \bibinfo
  {author} {\bibfnamefont {F.}~\bibnamefont {Nejadsattari}}, \bibinfo {author}
  {\bibfnamefont {Y.}~\bibnamefont {Gefen}}, \ and\ \bibinfo {author}
  {\bibfnamefont {E.}~\bibnamefont {Karimi}},\ }\bibfield  {title} {\enquote
  {\bibinfo {title} {Geometric phase from {Aharonov}{\textendash}{Bohm} to
  {Pancharatnam}{\textendash}{Berry} and~beyond},}\ }\href {\doibase
  10.1038/s42254-019-0071-1} {\bibfield  {journal} {\bibinfo  {journal} {Nat.
  Rev. Phys.}\ }\textbf {\bibinfo {volume} {1}},\ \bibinfo {pages} {437}
  (\bibinfo {year} {2019})}\BibitemShut {NoStop}%
\bibitem [{\citenamefont {Bliokh}\ \emph {et~al.}(2019)\citenamefont {Bliokh},
  \citenamefont {Alonso},\ and\ \citenamefont {Dennis}}]{2019Bliokh122401}%
  \BibitemOpen
  \bibfield  {author} {\bibinfo {author} {\bibfnamefont {K.~Y.}\ \bibnamefont
  {Bliokh}}, \bibinfo {author} {\bibfnamefont {M.~A.}\ \bibnamefont {Alonso}},
  \ and\ \bibinfo {author} {\bibfnamefont {M.~R.}\ \bibnamefont {Dennis}},\
  }\bibfield  {title} {\enquote {\bibinfo {title} {Geometric phases in {2D and
  3D} polarized fields: geometrical, dynamical, and topological aspects},}\
  }\href {\doibase 10.1088/1361-6633/ab4415} {\bibfield  {journal} {\bibinfo
  {journal} {Rep. Prog. Phys.}\ }\textbf {\bibinfo {volume} {82}},\ \bibinfo
  {pages} {122401} (\bibinfo {year} {2019})}\BibitemShut {NoStop}%
\bibitem [{\citenamefont {Zhao}\ \emph {et~al.}(2021)\citenamefont {Zhao},
  \citenamefont {Xu},\ and\ \citenamefont {Tong}}]{2021Zhao1935}%
  \BibitemOpen
  \bibfield  {author} {\bibinfo {author} {\bibfnamefont {P.~Z.}\ \bibnamefont
  {Zhao}}, \bibinfo {author} {\bibfnamefont {G.~F.}\ \bibnamefont {Xu}}, \ and\
  \bibinfo {author} {\bibfnamefont {D.~M.}\ \bibnamefont {Tong}},\ }\bibfield
  {title} {\enquote {\bibinfo {title} {Advances in nonadiabatic holonomic
  quantum computation},}\ }\href {\doibase 10.1360/TB-2021-0036} {\bibfield
  {journal} {\bibinfo  {journal} {Chinese Science Bulletin}\ }\textbf {\bibinfo
  {volume} {66}},\ \bibinfo {pages} {1935} (\bibinfo {year}
  {2021})}\BibitemShut {NoStop}%
\bibitem [{\citenamefont {Bender}\ and\ \citenamefont
  {Papanicolaou}(1988)}]{1988Bender561}%
  \BibitemOpen
  \bibfield  {author} {\bibinfo {author} {\bibfnamefont {C.~M.}\ \bibnamefont
  {Bender}}\ and\ \bibinfo {author} {\bibfnamefont {N.}~\bibnamefont
  {Papanicolaou}},\ }\bibfield  {title} {\enquote {\bibinfo {title} {{WKB}
  calculation of quantum adiabatic phases and nonadiabatic corrections},}\
  }\href {\doibase 10.1051/jphys:01988004904056100} {\bibfield  {journal}
  {\bibinfo  {journal} {J. Phys. France}\ }\textbf {\bibinfo {volume} {49}},\
  \bibinfo {pages} {561} (\bibinfo {year} {1988})}\BibitemShut {NoStop}%
\bibitem [{\citenamefont {Wang}(1990)}]{1990Wang5107}%
  \BibitemOpen
  \bibfield  {author} {\bibinfo {author} {\bibfnamefont {S.-J.}\ \bibnamefont
  {Wang}},\ }\bibfield  {title} {\enquote {\bibinfo {title} {Nonadiabatic
  {Berry}'s phase for a spin particle in a rotating magnetic field},}\ }\href
  {\doibase 10.1103/PhysRevA.42.5107} {\bibfield  {journal} {\bibinfo
  {journal} {Phys. Rev. A}\ }\textbf {\bibinfo {volume} {42}},\ \bibinfo
  {pages} {5107} (\bibinfo {year} {1990})}\BibitemShut {NoStop}%
\bibitem [{\citenamefont {{N. Guang-jiong and S.-q. Chen and Y.-l.
  Shen}}(1995)}]{1995GCS100}%
  \BibitemOpen
  \bibfield  {author} {\bibinfo {author} {\bibnamefont {{N. Guang-jiong and
  S.-q. Chen and Y.-l. Shen}}},\ }\bibfield  {title} {\enquote {\bibinfo
  {title} {Geometric phase in spin precession and the adiabatic
  approximation},}\ }\href {\doibase 10.1016/0375-9601(94)00929-J} {\bibfield
  {journal} {\bibinfo  {journal} {Phys. Lett. A}\ }\textbf {\bibinfo {volume}
  {197}},\ \bibinfo {pages} {100} (\bibinfo {year} {1995})}\BibitemShut
  {NoStop}%
\bibitem [{\citenamefont {Tong}\ \emph {et~al.}(2005)\citenamefont {Tong},
  \citenamefont {Singh}, \citenamefont {Kwek}, \citenamefont {Fan},\ and\
  \citenamefont {Oh}}]{2005Tong288}%
  \BibitemOpen
  \bibfield  {author} {\bibinfo {author} {\bibfnamefont {D.~M.}\ \bibnamefont
  {Tong}}, \bibinfo {author} {\bibfnamefont {K.}~\bibnamefont {Singh}},
  \bibinfo {author} {\bibfnamefont {L.~C.}\ \bibnamefont {Kwek}}, \bibinfo
  {author} {\bibfnamefont {X.~J.}\ \bibnamefont {Fan}}, \ and\ \bibinfo
  {author} {\bibfnamefont {C.~H.}\ \bibnamefont {Oh}},\ }\bibfield  {title}
  {\enquote {\bibinfo {title} {A note on the geometric phase in adiabatic
  approximation},}\ }\href {\doibase 10.1016/j.physleta.2005.03.043} {\bibfield
   {journal} {\bibinfo  {journal} {Phys. Lett. A}\ }\textbf {\bibinfo {volume}
  {339}},\ \bibinfo {pages} {288} (\bibinfo {year} {2005})}\BibitemShut
  {NoStop}%
\bibitem [{\citenamefont {Chiara}\ and\ \citenamefont
  {Palma}(2003)}]{2003Chiara90404}%
  \BibitemOpen
  \bibfield  {author} {\bibinfo {author} {\bibfnamefont {G.~D.}\ \bibnamefont
  {Chiara}}\ and\ \bibinfo {author} {\bibfnamefont {G.~M.}\ \bibnamefont
  {Palma}},\ }\bibfield  {title} {\enquote {\bibinfo {title} {Berry phase for a
  spin $1/2$ particle in a classical fluctuating field},}\ }\href {\doibase
  10.1103/PhysRevLett.91.090404} {\bibfield  {journal} {\bibinfo  {journal}
  {Phys. Rev. Lett.}\ }\textbf {\bibinfo {volume} {91}},\ \bibinfo {pages}
  {090404} (\bibinfo {year} {2003})}\BibitemShut {NoStop}%
\bibitem [{\citenamefont {Carollo}\ \emph {et~al.}(2003)\citenamefont
  {Carollo}, \citenamefont {Fuentes-Guridi}, \citenamefont {Santos},\ and\
  \citenamefont {Vedral}}]{2003Carollo160402}%
  \BibitemOpen
  \bibfield  {author} {\bibinfo {author} {\bibfnamefont {A.}~\bibnamefont
  {Carollo}}, \bibinfo {author} {\bibfnamefont {I.}~\bibnamefont
  {Fuentes-Guridi}}, \bibinfo {author} {\bibfnamefont {M.~F.}\ \bibnamefont
  {Santos}}, \ and\ \bibinfo {author} {\bibfnamefont {V.}~\bibnamefont
  {Vedral}},\ }\bibfield  {title} {\enquote {\bibinfo {title} {Geometric phase
  in open systems},}\ }\href {\doibase 10.1103/PhysRevLett.90.160402}
  {\bibfield  {journal} {\bibinfo  {journal} {Phys. Rev. Lett.}\ }\textbf
  {\bibinfo {volume} {90}},\ \bibinfo {pages} {160402} (\bibinfo {year}
  {2003})}\BibitemShut {NoStop}%
\bibitem [{\citenamefont {Solinas}\ \emph {et~al.}(2004)\citenamefont
  {Solinas}, \citenamefont {Zanardi},\ and\ \citenamefont
  {Zangh{\`{\i}}}}]{2004Solinas42316}%
  \BibitemOpen
  \bibfield  {author} {\bibinfo {author} {\bibfnamefont {P.}~\bibnamefont
  {Solinas}}, \bibinfo {author} {\bibfnamefont {P.}~\bibnamefont {Zanardi}}, \
  and\ \bibinfo {author} {\bibfnamefont {N.}~\bibnamefont {Zangh{\`{\i}}}},\
  }\bibfield  {title} {\enquote {\bibinfo {title} {Robustness of non-abelian
  holonomic quantum gates against parametric noise},}\ }\href {\doibase
  10.1103/PhysRevA.70.042316} {\bibfield  {journal} {\bibinfo  {journal} {Phys.
  Rev. A}\ }\textbf {\bibinfo {volume} {70}},\ \bibinfo {pages} {042316}
  (\bibinfo {year} {2004})}\BibitemShut {NoStop}%
\bibitem [{\citenamefont {Zhu}\ and\ \citenamefont
  {Zanardi}(2005)}]{2005Zhu20301}%
  \BibitemOpen
  \bibfield  {author} {\bibinfo {author} {\bibfnamefont {S.-L.}\ \bibnamefont
  {Zhu}}\ and\ \bibinfo {author} {\bibfnamefont {P.}~\bibnamefont {Zanardi}},\
  }\bibfield  {title} {\enquote {\bibinfo {title} {Geometric quantum gates that
  are robust against stochastic control errors},}\ }\href {\doibase
  10.1103/PhysRevA.72.020301} {\bibfield  {journal} {\bibinfo  {journal} {Phys.
  Rev. A}\ }\textbf {\bibinfo {volume} {72}},\ \bibinfo {pages} {020301(R)}
  (\bibinfo {year} {2005})}\BibitemShut {NoStop}%
\bibitem [{\citenamefont {Fuentes-Guridi}\ \emph {et~al.}(2005)\citenamefont
  {Fuentes-Guridi}, \citenamefont {Girelli},\ and\ \citenamefont
  {Livine}}]{2005FuentesGuridi20503}%
  \BibitemOpen
  \bibfield  {author} {\bibinfo {author} {\bibfnamefont {I.}~\bibnamefont
  {Fuentes-Guridi}}, \bibinfo {author} {\bibfnamefont {F.}~\bibnamefont
  {Girelli}}, \ and\ \bibinfo {author} {\bibfnamefont {E.}~\bibnamefont
  {Livine}},\ }\bibfield  {title} {\enquote {\bibinfo {title} {Holonomic
  quantum computation in the presence of decoherence},}\ }\href {\doibase
  10.1103/PhysRevLett.94.020503} {\bibfield  {journal} {\bibinfo  {journal}
  {Phys. Rev. Lett.}\ }\textbf {\bibinfo {volume} {94}},\ \bibinfo {pages}
  {020503} (\bibinfo {year} {2005})}\BibitemShut {NoStop}%
\bibitem [{\citenamefont {Florio}\ \emph {et~al.}(2006)\citenamefont {Florio},
  \citenamefont {Facchi}, \citenamefont {Fazio}, \citenamefont {Giovannetti},\
  and\ \citenamefont {Pascazio}}]{2006Florio22327}%
  \BibitemOpen
  \bibfield  {author} {\bibinfo {author} {\bibfnamefont {G.}~\bibnamefont
  {Florio}}, \bibinfo {author} {\bibfnamefont {P.}~\bibnamefont {Facchi}},
  \bibinfo {author} {\bibfnamefont {R.}~\bibnamefont {Fazio}}, \bibinfo
  {author} {\bibfnamefont {V.}~\bibnamefont {Giovannetti}}, \ and\ \bibinfo
  {author} {\bibfnamefont {S.}~\bibnamefont {Pascazio}},\ }\bibfield  {title}
  {\enquote {\bibinfo {title} {Robust gates for holonomic quantum
  computation},}\ }\href {\doibase 10.1103/PhysRevA.73.022327} {\bibfield
  {journal} {\bibinfo  {journal} {Phys. Rev. A}\ }\textbf {\bibinfo {volume}
  {73}},\ \bibinfo {pages} {022327} (\bibinfo {year} {2006})}\BibitemShut
  {NoStop}%
\bibitem [{\citenamefont {Thomas}\ \emph {et~al.}(2011)\citenamefont {Thomas},
  \citenamefont {Lababidi},\ and\ \citenamefont {Tian}}]{2011Thomas42335}%
  \BibitemOpen
  \bibfield  {author} {\bibinfo {author} {\bibfnamefont {J.~T.}\ \bibnamefont
  {Thomas}}, \bibinfo {author} {\bibfnamefont {M.}~\bibnamefont {Lababidi}}, \
  and\ \bibinfo {author} {\bibfnamefont {M.}~\bibnamefont {Tian}},\ }\bibfield
  {title} {\enquote {\bibinfo {title} {Robustness of single-qubit geometric
  gate against systematic error},}\ }\href {\doibase
  10.1103/PhysRevA.84.042335} {\bibfield  {journal} {\bibinfo  {journal} {Phys.
  Rev. A}\ }\textbf {\bibinfo {volume} {84}},\ \bibinfo {pages} {042335}
  (\bibinfo {year} {2011})}\BibitemShut {NoStop}%
\bibitem [{SM()}]{SM}%
  \BibitemOpen
  \href@noop {} {}\bibinfo {note} {See Supplemental Material at [URL will be
  inserted by publisher] for the proof of
  Eq.~(\ref{GP-imperfect}).}\BibitemShut {Stop}%
\bibitem [{\citenamefont {Marzlin}\ and\ \citenamefont
  {Sanders}(2004)}]{2004Marzlin160408}%
  \BibitemOpen
  \bibfield  {author} {\bibinfo {author} {\bibfnamefont {K.-P.}\ \bibnamefont
  {Marzlin}}\ and\ \bibinfo {author} {\bibfnamefont {B.~C.}\ \bibnamefont
  {Sanders}},\ }\bibfield  {title} {\enquote {\bibinfo {title} {Inconsistency
  in the application of the adiabatic theorem},}\ }\href {\doibase
  10.1103/PhysRevLett.93.160408} {\bibfield  {journal} {\bibinfo  {journal}
  {Phys. Rev. Lett.}\ }\textbf {\bibinfo {volume} {93}},\ \bibinfo {pages}
  {160408} (\bibinfo {year} {2004})}\BibitemShut {NoStop}%
\bibitem [{\citenamefont {Jones}\ \emph {et~al.}(2000)\citenamefont {Jones},
  \citenamefont {Vedral}, \citenamefont {Ekert},\ and\ \citenamefont
  {Castagnoli}}]{2000Jones869}%
  \BibitemOpen
  \bibfield  {author} {\bibinfo {author} {\bibfnamefont {J.~A.}\ \bibnamefont
  {Jones}}, \bibinfo {author} {\bibfnamefont {V.}~\bibnamefont {Vedral}},
  \bibinfo {author} {\bibfnamefont {A.}~\bibnamefont {Ekert}}, \ and\ \bibinfo
  {author} {\bibfnamefont {G.}~\bibnamefont {Castagnoli}},\ }\bibfield  {title}
  {\enquote {\bibinfo {title} {Geometric quantum computation using nuclear
  magnetic resonance},}\ }\href {https://doi.org/10.1038/35002528} {\bibfield
  {journal} {\bibinfo  {journal} {Nature (London)}\ }\textbf {\bibinfo {volume}
  {403}},\ \bibinfo {pages} {869} (\bibinfo {year} {2000})}\BibitemShut
  {NoStop}%
\bibitem [{\citenamefont {Duan}\ \emph {et~al.}(2001)\citenamefont {Duan},
  \citenamefont {Cirac},\ and\ \citenamefont {Zoller}}]{2001Duan1695}%
  \BibitemOpen
  \bibfield  {author} {\bibinfo {author} {\bibfnamefont {L.-M.}\ \bibnamefont
  {Duan}}, \bibinfo {author} {\bibfnamefont {J.~I.}\ \bibnamefont {Cirac}}, \
  and\ \bibinfo {author} {\bibfnamefont {P.}~\bibnamefont {Zoller}},\
  }\bibfield  {title} {\enquote {\bibinfo {title} {Geometric manipulation of
  trapped ions for quantum computation},}\ }\href {\doibase
  10.1126/science.1058835} {\bibfield  {journal} {\bibinfo  {journal}
  {Science}\ }\textbf {\bibinfo {volume} {292}},\ \bibinfo {pages} {1695}
  (\bibinfo {year} {2001})}\BibitemShut {NoStop}%
\bibitem [{\citenamefont {Zu}\ \emph {et~al.}(2014)\citenamefont {Zu},
  \citenamefont {Wang}, \citenamefont {He}, \citenamefont {Zhang},
  \citenamefont {Dai}, \citenamefont {Wang},\ and\ \citenamefont
  {Duan}}]{2014Zu72}%
  \BibitemOpen
  \bibfield  {author} {\bibinfo {author} {\bibfnamefont {C.}~\bibnamefont
  {Zu}}, \bibinfo {author} {\bibfnamefont {W.-B.}\ \bibnamefont {Wang}},
  \bibinfo {author} {\bibfnamefont {L.}~\bibnamefont {He}}, \bibinfo {author}
  {\bibfnamefont {W.-G.}\ \bibnamefont {Zhang}}, \bibinfo {author}
  {\bibfnamefont {C.-Y.}\ \bibnamefont {Dai}}, \bibinfo {author} {\bibfnamefont
  {F.}~\bibnamefont {Wang}}, \ and\ \bibinfo {author} {\bibfnamefont {L.-M.}\
  \bibnamefont {Duan}},\ }\bibfield  {title} {\enquote {\bibinfo {title}
  {Experimental realization of universal geometric quantum gates with
  solid-state spins},}\ }\href {\doibase 10.1038/nature13729} {\bibfield
  {journal} {\bibinfo  {journal} {Nature}\ }\textbf {\bibinfo {volume} {514}},\
  \bibinfo {pages} {72} (\bibinfo {year} {2014})}\BibitemShut {NoStop}%
\bibitem [{\citenamefont {Farhi}\ \emph {et~al.}(2001)\citenamefont {Farhi},
  \citenamefont {Goldstone}, \citenamefont {Gutmann}, \citenamefont {Lapan},
  \citenamefont {Lundgren},\ and\ \citenamefont {Preda}}]{2001Farhi472}%
  \BibitemOpen
  \bibfield  {author} {\bibinfo {author} {\bibfnamefont {E.}~\bibnamefont
  {Farhi}}, \bibinfo {author} {\bibfnamefont {J.}~\bibnamefont {Goldstone}},
  \bibinfo {author} {\bibfnamefont {S.}~\bibnamefont {Gutmann}}, \bibinfo
  {author} {\bibfnamefont {J.}~\bibnamefont {Lapan}}, \bibinfo {author}
  {\bibfnamefont {A.}~\bibnamefont {Lundgren}}, \ and\ \bibinfo {author}
  {\bibfnamefont {D.}~\bibnamefont {Preda}},\ }\bibfield  {title} {\enquote
  {\bibinfo {title} {A quantum adiabatic evolution algorithm applied to random
  instances of an {NP}-complete problem},}\ }\href {\doibase
  10.1126/science.1057726} {\bibfield  {journal} {\bibinfo  {journal}
  {Science}\ }\textbf {\bibinfo {volume} {292}},\ \bibinfo {pages} {472}
  (\bibinfo {year} {2001})}\BibitemShut {NoStop}%
\bibitem [{\citenamefont {Zhang}\ \emph {et~al.}(2014)\citenamefont {Zhang},
  \citenamefont {Yu},\ and\ \citenamefont {Tong}}]{2014Zhang42321}%
  \BibitemOpen
  \bibfield  {author} {\bibinfo {author} {\bibfnamefont {D.-J.}\ \bibnamefont
  {Zhang}}, \bibinfo {author} {\bibfnamefont {X.-D.}\ \bibnamefont {Yu}}, \
  and\ \bibinfo {author} {\bibfnamefont {D.~M.}\ \bibnamefont {Tong}},\
  }\bibfield  {title} {\enquote {\bibinfo {title} {Theorem on the existence of
  a nonzero energy gap in adiabatic quantum computation},}\ }\href {\doibase
  10.1103/PhysRevA.90.042321} {\bibfield  {journal} {\bibinfo  {journal} {Phys.
  Rev. A}\ }\textbf {\bibinfo {volume} {90}},\ \bibinfo {pages} {042321}
  (\bibinfo {year} {2014})}\BibitemShut {NoStop}%
\bibitem [{\citenamefont {Zhu}\ \emph {et~al.}()\citenamefont {Zhu},
  \citenamefont {Lu},\ and\ \citenamefont {Lein}}]{Zhu}%
  \BibitemOpen
  \bibfield  {author} {\bibinfo {author} {\bibfnamefont {X.}~\bibnamefont
  {Zhu}}, \bibinfo {author} {\bibfnamefont {P.}~\bibnamefont {Lu}}, \ and\
  \bibinfo {author} {\bibfnamefont {M.}~\bibnamefont {Lein}},\ }\bibfield
  {title} {\enquote {\bibinfo {title} {Control of the geometric phase with
  time-dependent fields},}\ }\href
  {https://journals.aps.org/prl/accepted/29074Y58W4511c7a80331b17c55f7fca48e58b6e5}
  {\bibinfo  {journal} {Accepted by Phys. Rev. Lett.}\ }\BibitemShut {NoStop}%
\end{thebibliography}
%

\clearpage

\renewcommand{\theequation}{S\arabic{equation}}
\renewcommand{\thefigure}{S\arabic{figure}}

\def\eqkey{{14}}

\def\eqi{{9}}

\def\eqe{{11}}

\def\eqdef{{2}}

\setcounter{equation}{0}

\section*{\large{Supplemental Material}}

\begin{figure}
\includegraphics[width=0.4\textwidth]{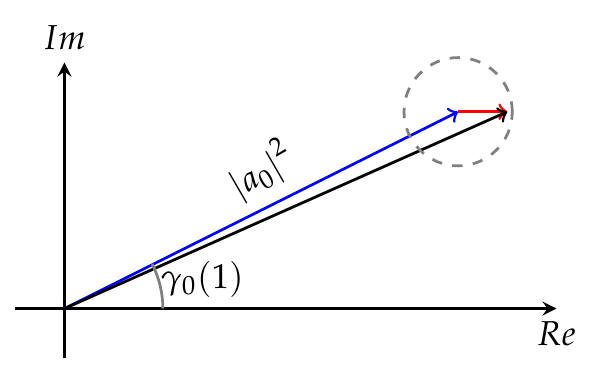}
\caption{Schematic representations of the three complex numbers $\sum_{n=0}^{N-1}\abs{a_n}^2e^{-iT\Delta_{n0}(1)}e^{i\gamma_n(1)}$ (the black line), $\abs{a_0}^2e^{i\gamma_0(1)}$ (the blue line), and $\sum_{n\neq 0}\abs{a_n}^2e^{-iT\Delta_{n0}(1)}e^{i\gamma_n(1)}$ (the red line) in the complex plane. Here, the red line is confined in the (gray dotted) circle centered at $\abs{a_0}^2e^{i\gamma_0(1)}$ and of radius $1-\abs{a_0}^2$.}
\label{SM-fig1}
\end{figure}

Here we present a proof of formula (\eqkey) in the main text. It follows from Eqs.~(\eqi) and (\eqe) in the main text that
\begin{eqnarray}\label{st1}
&&\arg\innerproduct{\psi_T^\prime(0)}{\psi_T^\prime(1)}
=\arg\sum_{n=0}^{N-1}
\abs{a_n}^2e^{-iT\int_0^1\varepsilon_n(s)ds}e^{i\gamma_n(1)}\nonumber\\
&&=\arg \left[e^{-iT\int_0^1\varepsilon_0(s)ds}
\sum_{n=0}^{N-1}\abs{a_n}^2e^{-iT\Delta_{n0}(1)}e^{i\gamma_n(1)}\right],
\end{eqnarray}
where
\begin{eqnarray}
\Delta_{mn}(s)=\int_0^s\left[\varepsilon_m(\sigma)-\varepsilon_n(\sigma)\right]d\sigma.
\end{eqnarray}
It is easy to see that
\begin{eqnarray}\label{st2}
\arg e^{-iT\int_0^1\varepsilon_0(s)ds}=-T\int_0^1\varepsilon_0(s)ds
~(\textrm{mod}~2\pi).
\end{eqnarray}
Here, for the sake of mathematical rigor, we invoke the notation $(\textrm{mod}~2\pi)$ to indicate that we work in modular arithmetic.
Besides, note that
\begin{eqnarray}\label{eq4}
&&\sum_{n=0}^{N-1}\abs{a_n}^2e^{-iT\Delta_{n0}(1)}e^{i\gamma_n(1)}=
\nonumber\\
&&\abs{a_0}^2e^{i\gamma_0(1)}
+\sum_{n\neq 0}\abs{a_n}^2e^{-iT\Delta_{n0}(1)}e^{i\gamma_n(1)}.
\end{eqnarray}
The three complex numbers appearing in Eq.~(\ref{eq4}) can be schematically represented in the complex plane as in Fig.~\ref{SM-fig1},
from which we deduce that
\begin{eqnarray}\label{st3}
\arg\sum_{n=0}^{N-1}\abs{a_n}^2e^{-iT\Delta_{n0}(1)}e^{i\gamma_n(1)}\approx\gamma_0(1)~
(\textrm{mod}~2\pi).
\end{eqnarray}
Substituting Eqs.~(\ref{st2}) and (\ref{st3}) into Eq.~(\ref{st1}) and noting that
\begin{eqnarray}
\arg (z_1 z_2)=\arg z_1+\arg z_2~(\textrm{mod}~2\pi),
\end{eqnarray}
for two complex numbers $z_1$ and $z_2$, we have
\begin{eqnarray}\label{eq:cal-3}
\arg\innerproduct{\psi_T^\prime(0)}{\psi_T^\prime(1)}
\approx\gamma_0(1)-T\int_0^1\varepsilon_0(s)ds
~
(\textrm{mod}~2\pi).\nonumber\\
\end{eqnarray}
On the other hand, a direct calculation shows that
\begin{eqnarray}\label{eq:cal-1}
&&i
\int_{0}^{1}\innerproduct{\psi_T^\prime(s)}
{\partial_s\psi_T^\prime(s)}ds=
\sum_{n=0}^{N-1}\abs{a_n}^2T\int_{0}^1\varepsilon_n(s)ds+\nonumber\\
&&i\sum_{m\neq n}a_m^*a_n\int_0^1 e^{iT\Delta_{mn}(s)}e^{i\left[\gamma_n(s)-\gamma_m(s)\right]}
\innerproduct{\varepsilon_m(s)}{\partial_s\varepsilon_n(s)}ds.\nonumber\\
\end{eqnarray}
Note that $e^{{iT\Delta_{mn}(s)}}$ is a rapidly oscillating term for $m\neq n$, which implies that the integral
\begin{eqnarray}
\int_0^1 e^{iT\Delta_{mn}(s)}e^{i\left[\gamma_n(s)-\gamma_m(s)\right]}
\innerproduct{\varepsilon_m(s)}{\partial_s\varepsilon_n(s)}ds\rightarrow 0,\nonumber\\
\end{eqnarray}
as $T\rightarrow \infty$. Besides, $\abs{a_m^*a_n}\ll 1$ for $m\neq n$. Hence, the second term on the right-hand side of Eq.~(\ref{eq:cal-1}) is very small, and we have
\begin{eqnarray}\label{eq:cal-2}
i
\int_{0}^{1}\innerproduct{\psi_T^\prime(s)}
{\partial_s\psi_T^\prime(s)}ds\approx
\sum_{n=0}^{N-1}\abs{a_n}^2T\int_{0}^1\varepsilon_n(s)ds.\nonumber\\
\end{eqnarray}
Inserting Eqs.~(\ref{eq:cal-3}) and (\ref{eq:cal-2}) into Eq.~(\eqdef) in the main text, we obtain
\begin{eqnarray}\label{GP-imperfect}
\gamma[\mathcal{P}_T^\prime]\approx\gamma_0(1)+\sum_{n\neq 0}\abs{a_n}^2\Delta_{n0}(1)T
~
(\textrm{mod}~2\pi),
\end{eqnarray}
i.e., formula (\eqkey) in the main text.

\end{document}